# 多様な移行元言語からの共通的GPU自動オフロード検討

山登庸次†

† NTT ネットワークサービスシステム研究所，東京都武蔵野市緑町 3-9-11
E-mail: †yoji.yamato.wa@hco.ntt.co.jp

**あらまし**　近年，少コア CPU だけでなく，GPU，FPGA，メニーコア CPU 等のヘテロなデバイスを利用したシステムが増えている．しかし，これらの利用には，CUDA 等のハードウェアを意識した技術仕様の理解が必要であり，ハードルは高い．これらの背景から，私は，プログラマーが CPU 向けに開発したソースコードを，適用される環境に応じて，自動で変換し，リソース量等を設定して，高い性能で運用可能とする環境適応ソフトウェアのコンセプトを提案している．しかし，従来，オフロードする移行元の言語は C 言語のアプリケーションが主流で，複数の言語のアプリケーションに対して共通的にオフロードするための研究は無かった．本稿では，C 言語だけでなく，Python，Java と移行元言語が多様となった場合でも，アプリケーションを自動でオフロードするための共通的方式を検討する．

# Study of Automatic GPU Offloading Method from Various Language Applications

Yoji YAMATO†

† Network Service Systems Laboratories, NTT Corporation, 3-9-11, Midori-cho, Musashino-shi, Tokyo
E-mail: †yoji.yamato.wa@hco.ntt.co.jp

**Abstract**　In recent years, utilization of heterogeneous hardware other than small core CPU such as GPU, FPGA or many core CPU is increasing. However, when using heterogeneous hardware, barriers of technical skills such as CUDA are high. Based on that, I have proposed environment-adaptive software that enables automatic conversion, configuration, and high performance operation of once written code, according to the hardware to be placed. However, the source language for offloading was mainly C/C++ language applications currently, and there was no research for common offloading for various language applications. In this paper, I study a common method for automatically offloading for various language applications not only in C language but also in Python and Java.

## 1. は じ め に

近年，CPU だけでなく，FPGA（Field Programmable Gate Array）や GPU（Graphics Processing Unit）等のデバイスの活用が増えている．例えば，Microsoft 社は FPGA を使って Bing の検索効率を高めるといった取り組みをしており [1]，Amazon 社は，FPGA, GPU 等をクラウドのインスタンス（例えば，[2]- [10]）として提供している [11]．また，IoT デバイスを用いた IoT システム（例えば，[12]- [16]）も，サービス連携技術等（例えば，[17]- [25]）を用いて開発され増えてきている．

しかし，少コアの CPU 以外のデバイスをシステムで適切に活用するためには，デバイス特性を意識した設定やプログラム作成が必要であり，OpenMP（Open Multi-Processing）[26]，OpenCL（Open Computing Language）[27]，CUDA（Compute Unified Device Architecture）[28] といった知識が必要になるため，大半のプログラマーにとっては，スキルの壁が高い．

そこで，そのような壁を取り払い，少コアの CPU 以外のデバイスを十分利用できるようにするため，プログラマーが処理ロジックを記述したソフトウェアを，配置先の環境（FPGA，GPU，メニーコア CPU 等）にあわせて，適応的に変換，設定し，環境に適合した動作をさせることが求められている．

私は，一度記述したコードを，配置先の環境に存在する GPU や FPGA，メニーコア CPU 等を利用できるように，変換，リソース設定等を自動で行い，アプリケーションを高性能に動作させることを目的とした，環境適応ソフトウェアを提案した．合わせて，環境適応ソフトウェアの要素として，C 言語プログ

ラムコードのループ文及び機能ブロックを，FPGA，GPU に自動オフロードする方式を提案評価している [29]．本稿では，C 言語だけでなく，Python，Java と移行元言語が多様となった場合でも，アプリケーションを自動でオフロードすることを目的とする．まず，基本的なフローはどの言語でも同じ共通的方式を提案する．次に，移行元言語が，C 言語，Python，Java の時の言語依存処理を検討する．

## 2. 既 存 技 術

GPGPU（General Purpose GPU）（例えば [30] [31]）を行うための環境として CUDA が普及している．CUDA は GPGPU 向けの NVIDIA 社の環境だが，FPGA，メニーコア CPU，GPU 等のヘテロなデバイスを同じように扱うための仕様として OpenCL が出ており，その開発環境 [32] [33] も出てきている．CUDA，OpenCL は，C 言語の拡張を行いプログラムを行う形だが，プログラムの難度は高い（カーネルとホストとの間のメモリデータのコピーや解放の記述を明示的に行う等）

より簡易にヘテロなデバイスを利用するため，指示行ベースで，並列処理等を行う箇所を指定して，指示行に従ってコンパイラが，GPU，メニーコア CPU 等に向けて実行ファイルを作成する技術がある．仕様としては，OpenACC [34] や OpenMP 等，コンパイラとして PGI コンパイラ [35] や gcc 等がある．

CUDA，OpenCL，OpenACC，OpenMP 等の技術仕様を用いることで，FPGA や GPU，メニーコア CPU へオフロードすることは可能になっている．しかしデバイス処理自体は行えるようになっても，高速化することには課題がある．例えば，マルチコア CPU 向けに自動並列化機能を持つコンパイラとして，Intel コンパイラ [36] 等がある．これらは，自動並列化時に，コードの中のループ文中で並列処理可能な部分を抽出して，並列化している．しかし，メモリ処理等の影響で単に並列化可能ループ文を並列化しても性能がでないことも多い．FPGA や GPU 等で高速化する際には，OpenCL や CUDA の技術者がチューニングを繰り返したり，OpenACC コンパイラ等を用いて適切な並列処理範囲を探索し試行することがされている．

このため，技術スキルが乏しいプログラマーが，FPGA や GPU，メニーコア CPU を活用してアプリケーションを高速化することは難しいし，自動並列化技術等を使う場合も並列処理箇所探索の試行錯誤等の稼働が必要だった．

現状，ヘテロなデバイスに対するオフロードは手動での取組みが主流である．また，著者は環境適応ソフトウェアのコンセプトを提案し（図 1），自動オフロードを検討しているが，C 言語プログラムの GPU や FPGA へのオフロードである等，Python，Java 等の多様な移行元言語は想定されていない．

## 3. 多様な移行元言語での自動オフロードの検討

### 3.1 多様移行元言語対応の基本的考え

本稿で対象とする多様な移行元言語としては，C 言語，Python，Java の 3 つとする．これら 3 つは，毎月 TIOBE が発表するプログラム言語の人気ランキングの上位 3 つであり，プログラマー人口が多い．また，C 言語はコンパイル型，Python はイ

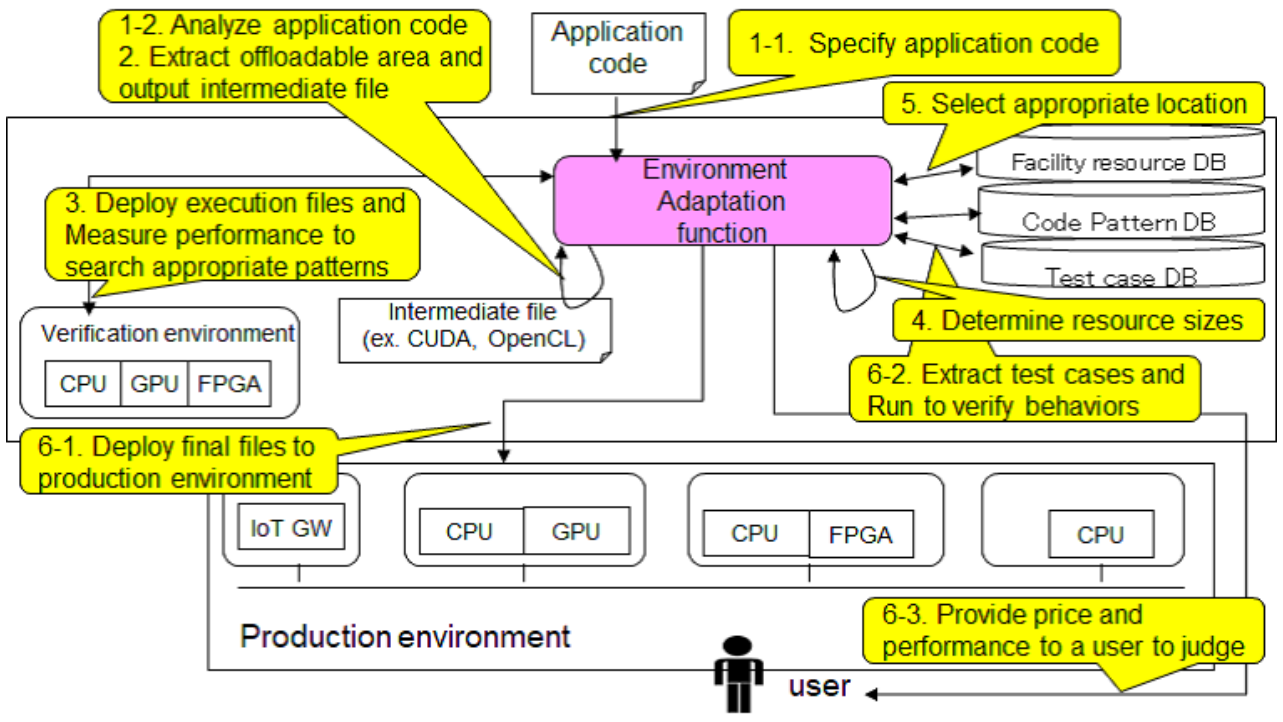

図 1 環境適応ソフトウェアのフロー

ンタプリタ型，Java はその中間的方式と，方式上の多様性も 3 つでカバーされている．そのため，これら 3 つで共通的に利用できる方式であれば，より多くの言語への対応も容易と考える．

移行先環境が単なる CPU でない場合で，多様な移行元言語プログラムを，自動で高速にオフロードするために，検証環境の実機で性能測定し，進化計算手法等の手法と組み合わせて，徐々に高速なオフロードパターンを見つけるアプローチをとる．理由として，性能に関しては，コード構造だけでなく，処理するハードウェアのスペック，コンパイラやインタプリタ，データサイズ，ループ回数等の処理内容によって大きく変わるため，静的に予測する事が困難であり，動的な測定が必要だからである．実際に，市中には，ループ文を見つけコンパイル段階で並列化する自動並列化コンパイラがあるが，並列化可能ループ文の並列化だけでは性能を測定してみると低速になる場合も多いため，性能測定は必要である．

また，オフロードする対象については，プログラムのループ文と機能ブロックとするアプローチをとる．ループ文については，処理時間がかかるプログラムの処理の大半はループで費やされているという現状から，ループ文がオフロードのターゲットとしてはまず考えられる．一方，機能ブロックについては，特定処理を高速化する際に，処理内容や処理ハードウェアに適したアルゴリズムを用いることが多いため，個々のループ文の並列処理等に比べ，大きく高速化できる場合がある．行列積算やフーリエ変換等の頻繁に使われる機能ブロック単位で，GPU 等の処理デバイスに応じたアルゴリズムで実装された処理（CUDA ライブラリ等）に置換することで高速化する．

移行先環境としては，著者はこれまで GPU，FPGA，メニーコア CPU の 3 つを考えてきており（[37]- [40]），これらが混在した環境での C 言語プログラムのオフロードも検討してきた．今回は移行元言語の多様化が主となるため，評価する移行先としては GPU の一つとする．FPGA やメニーコア CPU については，GPU で共通的方式を確認できれば，その拡張となる．

### 3.2 共通的な GPU オフロード手法

#### 3.2.1 ループ文の GPU 自動オフロード

ループ文の GPU 自動オフロード手法については，[29] [37] で提案している．

ループ文を GPU にオフロードすることでどの程度の性能になるかは，実測してみないと予測は困難である．そのため，この

ループを GPU にオフロードするという指示を手動で行い，性能測定を試行錯誤することが頻繁に行われている．[29] はそれを踏まえ，GPU にオフロードする適切なループ文の発見を，進化計算手法の一つである遺伝的アルゴリズム（Genetic Algorithms: GA）[41] で自動的に行うことを提案している．GPU 処理を想定していない通常 CPU 向け汎用プログラムから，最初に並列可能ループ文のチェックを行い，次に並列可能ループ文群に対して，GPU 実行の際を 1，CPU 実行の際を 0 と値を置いて遺伝子化し，検証環境で性能検証試行を反復し適切な領域を探索している．

[37] [38] では，ループ文の適切な抽出に加えて，ネストループ文の中で利用される変数について，ループ文を GPU にオフロードする際に，ネストの下位で CPU-GPU 転送が行われると下位のループの度に転送が行われ効率的でないため，上位で CPU-GPU 転送が行われても問題ない変数については，上位でまとめて転送を行うことを提案している．ループ文オフロードの共通的方式では，コードの分析，ループと変数の把握，ループの GPU 処理有無の遺伝子化，遺伝子情報のコード化，コンパイル，検証環境での性能測定，次世代遺伝子の作成，これら処理の反復実行，最終解の決定が行われる．

#### 3.2.2 機能ブロックの自動オフロード

機能ブロックの自動オフロード手法については，[40] で提案を行っている．機能ブロックオフロードの処理概要を説明する．Step1 にて，ソースコードの分析が行われるが，Clang 等の構文解析ツールを用いて，ループ文構造等とともに，コードに含まれるライブラリ呼び出しや，機能処理を分析する．Step1 で把握したライブラリ呼び出しや機能処理について，Step2 で，コードパターン DB と照合することで，GPU，FPGA にオフロードできる処理を発見する．Step3 で，オフロードできる処理について，GPU 向けのライブラリ等に置換して，CPU プログラムとのインタフェースを作成することでオフロードする．なお，機能処理に対して，高速化するライブラリや IP コアがあるかの探索は，ライブラリ等の名前一致に加えて類似性検出ツールでの検出も行う．類似性検出ツールとは，Deckard [42] や CCFinderX [43] 等の，コピーコードやコピー後変更したコードの検出行うツールである．

[40] で評価の通り，個々のループ文のオフロードに比べ，複数のループ文を含む機能ブロック単位で処理するデバイスにアルゴリズム含めてチューンした機能ブロックオフロードは高速化度合いが高い．

機能ブロックオフロードの共通的方式では，コードの分析，機能ブロックの把握，オフロード可能機能ブロックの探索，オフロード機能ブロックへの置換，コンパイル，検証環境での性能測定，これら処理の反復実行，最終解の決定が行われる．

### 3.3 移行元言語の対応検討

#### 3.3.1 C 言語への対応検討

基本的フローは共通的方式で言語非依存にできるが，個々の C 言語に依存・非依存の処理について記載する．

ループ文オフロードの，コードの分析では，C 言語を解析する Clang [44] 等の構文解析ツールを用いて構文解析する．ループと変数の把握については，構文解析ツールの結果を管理する際は，言語に非依存に抽象的に管理できる．ループの GPU 処理有無の遺伝子化についても，言語に非依存である．遺伝子情報のコード化では，遺伝子情報に合わせて GPU で実行するためのコードを作成するため，C 言語の拡張文法である OpenACC で GPU 処理を指定したり，変数転送を指定したりする．コンパイルは，OpenACC コードを PGI コンパイラ等でコンパイルする．性能測定は，言語に合わせて，Jenkins [45]、Selenium [46] 等の自動測定ツールも用いて行う．次世代の遺伝子作成は，性能測定結果に合わせて適合度を設定し交叉等の処理を行うが，言語に非依存である．反復実行と最終解の決定も，言語に非依存である．

機能ブロックオフロードの，コードの分析では，C 言語を解析する Clang 等の構文解析ツールを用いて構文解析する．機能ブロックの把握については，構文解析ツールの結果を用いて，次処理のマッチング探索に用いるため，言語に非依存の機能ブロックとして管理する．オフロード可能機能ブロックの探索では，ライブラリ等の名前一致でのマッチングと，Deckard 等の C 言語機能ブロックの類似性検出ツールを用いた類似性検知による，探索が行われる．オフロード機能ブロックへの置換は，CUDA ライブラリ呼び出し等，その言語からのオフロード機能利用に合わせた処理に置換する必要がある．コンパイルは，CUDA ライブラリ呼び出し等の C 言語コードを PGI コンパイラ等でコンパイルする．性能測定は，言語に合わせて，Jenkins 等の自動測定ツールも用いて行う．オフロード可能機能ブロックが複数の際は反復実行され，最高性能のパターンが最終解として決定される．

#### 3.3.2 Python への対応検討

ループ文オフロードの，コードの分析では，Python を解析する ast [47] 等の構文解析ツールを用いて構文解析する．ループと変数の把握については，構文解析ツールの結果を管理する際は，言語に非依存に抽象的に管理できる．ループの GPU 処理有無の遺伝子化についても，言語に非依存である．遺伝子情報のコード化では，遺伝子情報に合わせて GPU で実行するためのコードを作成するため，CUDA 文法で GPU 処理を指定したり，変数転送を指定したりする．インタプリタは，CUDA での指示を追加した Python コードを PyCUDA [48] でインタプリットする．性能測定は，言語に合わせて，Jenkins 等の自動測定ツールも用いて行う．次世代の遺伝子作成は，性能測定結果に合わせて適合度を設定し交叉等の処理を行うが，言語に非依存である．反復実行と最終解の決定も，言語に非依存である．

機能ブロックオフロードの，コードの分析では，Python を解析する ast 等の構文解析ツールを用いて構文解析する．機能ブロックの把握については，構文解析ツールの結果を用いて，次処理のマッチング探索に用いるため，言語に非依存の機能ブロックとして管理する．オフロード可能機能ブロックの探索では，ライブラリ等の名前一致でのマッチングと，CloneDigger 等の Python 機能ブロックの類似性検出ツールを用いた類似性検知による，探索が行われる．オフロード機能ブロックへの置換は，GPU 処理の pyCUDA での呼び出し等，その言語からの

オフロード機能利用に合わせた処理に置換する必要がある．インタプリタは，CUDA に合わせた Python コードを PyCUDA でインタプリットする．性能測定は，言語に合わせて，Jenkins 等の自動測定ツールも用いて行う．オフロード可能機能ブロックが複数の際は反復実行され，最高性能のパターンが最終解として決定される．

#### 3.3.3 Java への対応検討

ループ文オフロードの，コードの分析では，Java を解析する JavaParser 等の構文解析ツールを用いて構文解析する．ループと変数の把握については，構文解析ツールの結果を管理する際は，言語に非依存に抽象的に管理できる．ループの GPU 処理有無の遺伝子化についても，言語に非依存である．遺伝子情報のコード化では，遺伝子情報に合わせて GPU で実行するためのコードを作成するため，Java のラムダ記述で GPU 処理を指定したり，変数転送を指定したりする．実行環境は，Java のラムダ記述での並列化を GPU に対して行うことができる IBM JDK [49] を用いる．性能測定は，言語に合わせて，Jenkins 等の自動測定ツールも用いて行う．次世代の遺伝子作成は，性能測定結果に合わせて適合度を設定し交叉等の処理を行うが，言語に非依存である．反復実行と最終解の決定も，言語に非依存である．

機能ブロックオフロードの，コードの分析では，Java を解析する JavaParser 等の構文解析ツールを用いて構文解析する．機能ブロックの把握については，構文解析ツールの結果を用いて，次処理のマッチング探索に用いるため，言語に非依存の機能ブロックとして管理する．オフロード可能機能ブロックの探索では，ライブラリ等の名前一致でのマッチングと，Deckard 等の Java 機能ブロックの類似性検出ツールを用いた類似性検知による，探索が行われる．オフロード機能ブロックへの置換は，GPU 処理の CUDA ライブラリの呼び出し等，その言語からのオフロード機能利用に合わせた処理に置換する必要がある．実行環境は，Java のラムダ記述での処理を GPU に対して行うことができる IBM JDK を用いる．性能測定は，言語に合わせて，Jenkins 等の自動測定ツールも用いて行う．オフロード可能機能ブロックが複数の際は反復実行され，最高性能のパターンが最終解として決定される．

## 4. 実　　装

### 4.1 利用ツール

現在，Perl と Python で実装している実装の設計を説明する．

GPU 処理は，C/C++言語は PGI コンパイラ 19.10 を用いる．PGI コンパイラは OpenACC を解釈する C/C++向けコンパイラである．合わせて，cuFFT 等の CUDA ライブラリの呼び出しも処理が可能である．Python は PyCUDA 2019.1.2 を用いる．PyCUDA は Python から GPU に処理実行するためのインタプリタである．Java は IBM JDK を用いる．IBM JDK は Java のラムダ記述に従って並列処理を GPU に対して実行する仮想マシンである．

C/C++言語の構文解析には，LLVM/Clang 6.0 の構文解析ライブラリ (libClang の python binding) [?] を用いる．Python の構文解析には，ast を用いる．Java の構文解析には，Java Parser を用いる．

類似性検出ツールには，C/C++言語，Java には，Deckard v2.0 [42] を用いる．Python には，CloneDigger を用いる．照合に用いるコードパターン DB は，MySQL8 を用いる．ライブラリ等を類似性検出技術で検出するための，比較用コードとの対応関係等が保持される．

### 4.2 実装動作

実装の動作概要を示す．実装は，アプリケーションの利用依頼があると，構文解析ライブラリを用いてコード解析を行う．次に，機能ブロックオフロード，ループ文オフロードの順に試行を行う．これは，ループ文と機能ブロックに関しては，アルゴリズム含めて処理内容に合わせてオフロードする機能ブロックオフロードの方が高速化できるからである．機能ブロックオフロードが可能だった場合は，後半のループ文オフロードはオフロード可能だった機能ブロック部分を抜いたコードに対して試行する．性能測定の結果，最高性能のパターンを解とする．

#### 4.2.1 機能ブロックオフロード試行

実装は，まず，コード解析にて，呼び出されているライブラリや定義されているクラス，構造体等のプログラム構造を把握する．

実装は，次に，呼び出されているライブラリを高速化できる GPU 用ライブラリ等の検出を行う．呼び出されているライブラリをキーに，コードパターン DB に登録されているレコードから，高速化可能な実行ファイル等を取得する．高速化できる置換用機能が見つかったら，実装は，その実行用ファイルを作成する．GPU 用ライブラリの場合は，置換用ライブラリ（CUDA ライブラリ等）を呼び出すよう，元の部分は削除して置換記述する．置換記述が終わったら，PGI コンパイラ等でコンパイルする．

ライブラリ呼び出しの場合について記載したが，類似性検知を用いる場合も並行して処理がされる．実装は，類似性検出ツールを用いて，検出されたクラス，構造体等の部分コードと DB に登録された比較用コードの類似性検知を行い，閾値越えの機能ブロックと該当する GPU 用ライブラリを発見する．特に置換元のコードと置換するライブラリの引数や戻り値，型等のインタフェースが異なる場合は，オフロードを依頼したユーザに対して，置換先ライブラリに合わせて，インタフェースを変更してよいか確認し，確認後に実行用ファイルを作成する．

ここで，検証環境の GPU で性能測定できる実行用ファイルが作成される．機能ブロックオフロードについては，置換機能ブロック一つずつに対してオフロードするしないを性能測定し高速化できるか確認し，複数ある場合はその組み合わせ対しても検証する．

#### 4.2.2 ループ文オフロード試行

実装は，まず，コードを解析して，for 文を発見するとともに，for 文内で使われる変数データ，その変数の処理等の，プログラム構造を把握する．

並列処理自体が不可な for 文は排除する必要がある．各 for 文に対して，GPU で処理する指示挿入を試行し，エラーが出

る for 文は GA の対象外とする．ここで，エラーが出ないループ文の数が a の場合，a が遺伝子長となる．

次に，初期値として，指定個体数の遺伝子配列を準備する．遺伝子の各値は，0 と 1 をランダムに割当てて作成する．準備された遺伝子配列に応じて，遺伝子の値が 1 の場合は GPU 処理の指示をコードに挿入する．

次に，変数データの参照関係を元に，GPU 向けデータ転送を指示する．転送が必要なケースは，CPU プログラム側で設定，定義した変数と GPU プログラム側で参照する変数が重なる場合は，CPU から GPU への変数転送が必要であり，GPU プログラム側で設定した変数と CPU プログラム側で参照，設定，定義する変数が重なる場合は，GPU から CPU への変数転送が必要である．転送必要な変数について，GPU 処理開始前と終了後に一括転送すればよい変数については，複数ファイルで定義された変数を一括転送する指示を挿入する．

指示を挿入されたコードを，PGI コンパイラ等でコンパイルを行う．コンパイルした実行ファイルをデプロイし性能を測定する．性能測定では処理時間とともに，PGI コンパイラの PCAST 機能等を用いて並列処理した場合の計算結果が，元のコードと大きく差分がないかチェックし，許容外の場合は，処理時間を∞とする．

全個体に対して，性能測定後，処理時間に応じて，各個体の適合度を設定する．設定された適合度に応じて，残す個体の選択を行う．選択された個体に対して，交叉処理，突然変異処理，そのままコピー処理の GA 処理を行い，次世代の個体群を作成する．

次世代の個体に対して，指示挿入，コンパイル，性能測定，適合度設定，選択，交叉，突然変異処理を行う．指定世代数の GA 処理終了後，最高性能の遺伝子配列に該当する，指示付きコードを解とする．

### 4.3 共通機能部と多様な移行元言語向け処理

3 節の通り，機能ブロックオフロードでは，機能ブロックの管理と機能ブロックの名前一致でのマッチングについては共通機能となり，ループ文オフロードでは，ループと変数の管理と GA の遺伝子処理については共通機能となる．

#### 4.3.1 C 言語向け処理

ツール利用やコンパイラ向け指示設定は C 言語依存で実装する．構文解析には Clang を用いて，ループ文等の構造と機能ブロックについて把握する．機能ブロックのオフロードのため，Deckard を用いて置換機能探索が行われ，オフロードできる場合は，該当する GPU 向けライブラリを呼び出すようにして PGI コンパイラでコンパイルし，性能測定を Jenkins を用いて行う．

次に，ループ文のオフロードのため，ループパターンを遺伝子化して，OpenACC の#pragma acc kernels と#pragma acc paralell loop で GPU 処理を指定し，#pragma acc data copy と#pragma acc data present でデータ転送有無を指定する．遺伝子パターンに対応する OpenACC コードを PGI コンパイラでコンパイルし，性能測定を Jenkins を用いて行う．GA で次世代遺伝子パターンの作成，反復は共通機能処理として行う．

#### 4.3.2 Python 向け処理

構文解析には ast を用いて，ループ文等の構造と機能ブロックについて把握する．機能ブロックのオフロードのため，CloneDigger を用いて置換機能探索が行われ，オフロードできる場合は，該当する GPU 向けライブラリを呼び出すようにしてインタプリットし，性能測定を Jenkins を用いて行う．

次に，ループ文のオフロードのため，ループパターンを遺伝子化して，pyCUDA を用いて CUDA で GPU での計算処理とデータ転送有無を指定する．遺伝子パターンに対応する python CUDA コードをインタプリットし，性能測定を Jenkins を用いて行う．

#### 4.3.3 Java 向け処理

構文解析には JavaParser を用いて，ループ文等の構造と機能ブロックについて把握する．機能ブロックのオフロードのため，Deckard を用いて置換機能探索が行われ，オフロードできる場合は，該当する GPU 向けライブラリを呼び出すようにして IBM JDK で処理実行し，性能測定を Jenkins を用いて行う．

次に，ループ文のオフロードのため，ループパターンを遺伝子化して，Java ラムダ記述の java.util.Stream.IntStream.range(0, n).parallel().forEach で並列処理を指定する．遺伝子パターンに対応するラムダ記述の Java コードを IBM JDK で処理実行することで GPU にオフロードされるため，性能測定を Jenkins を用いて行う．

## 5. ま と め

本稿では，ソフトウェアを配置先環境に合わせて自動適応させ GPU，FPGA，メニーコア CPU 等を適切に利用して，アプリケーションを高性能に運用するための環境適応ソフトウェアの要素として，移行元言語が，C 言語，Python，Java と多様な際でも，自動オフロードできる方式を提案した．

移行元言語が何であっても，共通的に GPU に自動オフロードするための共通的方式を検討した．ループ文については，GPU にオフロードするに適したループ文オフロードパターンを進化的計算手法により自動探索し，各オフロードパターンに応じて不要な CPU-GPU 転送を削減するための転送指定を行う．機能ブロックについては，GPU にオフロード可能な機能ブロックを，ライブラリ呼び出しの名前一致や機能ブロックの類似性検出により発見し，オフロードする機能ブロックに置換する事で高速化する．次に，共通的方式を C 言語，Python，Java の各言語について処理する際に，言語に依存・非依存する処理を検討し，ループ，変数，機能ブロック管理や遺伝子処理を言語非依存に共通化するとともに，市中のツール，コンパイラ等を用いて言語依存処理を実装している．実装後，既存アプリケーションに対して，多様な言語での方式の有効性を示す．

## 文 献